# An Efficient Framework for Floor-plan Prediction of Dynamic Runtime Reconfigurable Systems

A. Al-Wattar, S. Areibi, G. Grewal
Faculty of Engineering and Computer Science
University of Guelph, Guelph, Canada

**ABSTRACT**

Several embedded application domains for reconfigurable systems tend to combine frequent changes with high performance demands of their workloads such as image processing, wearable computing and network processors. Time multiplexing of reconfigurable hardware resources raises a number of new issues, ranging from run-time systems to complex programming models that usually form a Reconfigurable hardware Operating System (ROS). The Operating System performs online task scheduling and handles resource management. There are many challenges in adaptive computing and dynamic reconfigurable systems. One of the major understudied challenges is estimating the required resources in terms of soft cores, Programmable Reconfigurable Regions (PRRs), the appropriate communication infrastructure, and to predict a near optimal layout and floor-plan of the reconfigurable logic fabric. Some of these issues are specific to the application being designed, while others are more general and relate to the underlying run-time environment. Static resource allocation for Run-Time Reconfiguration (RTR) often leads to inferior and unacceptable results. In this paper, we present a novel adaptive and dynamic methodology, based on a Machine Learning approach, for predicting and estimating the necessary resources for an application based on past historical information. An important feature of the proposed methodology is that the system is able to learn and generalize and, therefore, is expected to improve its accuracy over time. The goal of the entire process is to extract useful hidden knowledge from the data. This knowledge is the prediction and estimation of the necessary resources for an unknown or not previously seen application.

*Corresponding Author:*

Shawki Areibi
University of Guelph
50 Stone Road, Guelph, Ontario, Canada, 1-519-8244120
Email: sareibi@uoguelph.ca

## 1. INTRODUCTION

In the area of computer architecture choices span a wide spectrum, with Application Specific Integrated Circuits (ASICs) and General Purpose Processors (GPPs) being at opposite ends. General purpose processors are flexible, but unlike ASICs are not optimized to specific applications. Reconfigurable Architectures, in general, and Field Programmable Gate Arrays (FPGAs), in particular, fill the gap between these two extremes by achieving both the high performance of ASICs and the flexibility of GPPs. However, FPGAs are still not a match for the lower power consumed by ASICs nor the performance achieved by the latter . One important feature of FPGAs is their capability to adapt during the run-time of an application. The run-time reconfiguration environment capability in FPGAs provides common benefits in adapting hardware algorithms during system run-time, sharing hardware resources to reduce device count, power consumption, and shortening reconfiguration time [1].

Several embedded application domains for reconfigurable systems tend to combine frequent changes with high performance demands of their workloads such as image processing, wearable computing, and network processors. In many embedded systems used in different applications, several wireless standards and technologies, such as WiMax, WLAN, GSM, WCDMA have to be utilized and supported. However, it is unlikely that these protocols will be used simultaneously. Accordingly, it is possible to dynamically load only the one that is needed. Another example is employing different machine vision algorithms onto an Unmanned Ariel Vehicle (UAV) and utilizing the most appropriate based on the environment or perhaps the need to lower power consumption.

Time multiplexing of reconfigurable hardware resources raises a number of new issues, ranging from runtime systems to complex programming models that usually form a Reconfigurable hardware Operating System (ROS). The Operating System performs online task scheduling and handles resource management. The main objective of an Operating System for reconfigurable platforms is to reduce the complexity of applications development by giving the developer a higher level of abstraction with which to work.

### 1.1. Problem Definition

Performance is one of the fundamental reasons for using Reconfigurable Computing Systems (RCS). By mapping algorithms and applications to hardware, designers can tailor not only the computation components, but also perform data-flow optimization to match the algorithm. One of the main problems encountered in Run Time Reconfiguration (RTR) is identifying the most appropriate framework or infrastructure that suits an application. Typically, a designer would manually divide the FPGA fabric into static and dynamic regions [2]. The static regions would accommodate modules that do not change in time such as the task manager and necessary buses used for communication. The dynamic region is partitioned into a set of uniform or non-uniform modules with a certain size (we refer to these as "Programmable Reconfigurable Regions" (PRRs)) that act as application specific hardware accelerators for the incoming tasks that need to be executed. Every application (e.g., Machine Vision, Wireless-Sensor Network) requires specific types of resources that optimize certain objectives such as reducing power consumption, improving the execution time, reducing the cost or a combination of these objectives.

In current RTR systems, designers tend to perform resource allocation and floor-planning of the FPGA fabric a priori. These allocated resources, however, might not be appropriate for a new and different incoming application (e.g., streaming, non-streaming, hybrid). Instead of tailoring the FPGA fabric for an application, the latter has to suffer if the floor-plan is a miss match to the application itself. A one size fits all approach in this case would not only hinder the amount of performance sought by using RTR, but might even deteriorate speed, power, or the combination of both. As a result of this pre-specified, finite number of resource types and quantities allocated, a hefty price in terms of performance is often paid since the floor-plan and layout would not be suitable for the application in which it is being used for. This performance penalty may occur in the form of increased power consumption since meeting performance requirements might entail the usage of multiple PRRs. Accordingly an adaptive and dynamic approach is necessary for resource estimation and floorplanning.

### 1.2. Motivation

There are many challenges in adaptive computing and dynamic reconfigurable systems. One of the major understudied challenges is estimating and predicting the required resources in terms of soft cores, PRRs, and the appropriate communication infrastructure, to name just a few. Some of these issues are specific to the application being designed, while others are more general and relate to the underlying run-time environment. Static resource allocation for RTR might lead to inferior results. The number of PRRs for one application might be different than the number required by another. The type of PRRs (uniform, non-uniform, hybrid) also plays a crucial role in determining both performance and power consumption, as seen in Figure 1. The type of Scheduler used to determine when/where a task is executed is also important for specific real-time operations. The type of communication infrastructure that connects PRRs with the Task Manager plays an important role to speedup a certain application. Accordingly, in this work we seek to overcome the limitation of static resource allocation with a more appealing approach that can force the infrastructure of the reconfigurable computing platform to accommodate and match the application rather than the reverse.

### 1.3. Proposed Approach

In this paper, we present a novel adaptive and dynamic methodology based on an intelligent Machine Learning approach that is used to predict and estimate the necessary resources for an application based on past historical information. An important feature of the proposed methodology is that the system is able to learn as it gains more knowledge and, therefore, is expected to generalize and improve its accuracy over time. Even though the approach is general enough to predict most if not all types of resources from the number of PRRs, type of PRRs, the type of scheduler, and communication infrastructure, we limit our results to the former three required for an application. We plan to accomplish this task by first extracting certain features from the applications that are executed on the reconfigurable platform. The features compiled will be used to train and build a classification model that is capable of predicting the floorplan appropriate for an application. The classification model is a supervised learning approach that can generalize and accurately predict the class of an incoming application that it learned from previously seen patterns. Our proposed approach will be based on several modules including benchmark generation, data collection, pre-processing of data, data classification, and post processing. The goal of the entire process is to extract useful, hidden knowledge from the data, this knowledge is then used to predict and estimate the necessary resources and appropriate floorplan for an unknown or not previously seen application.

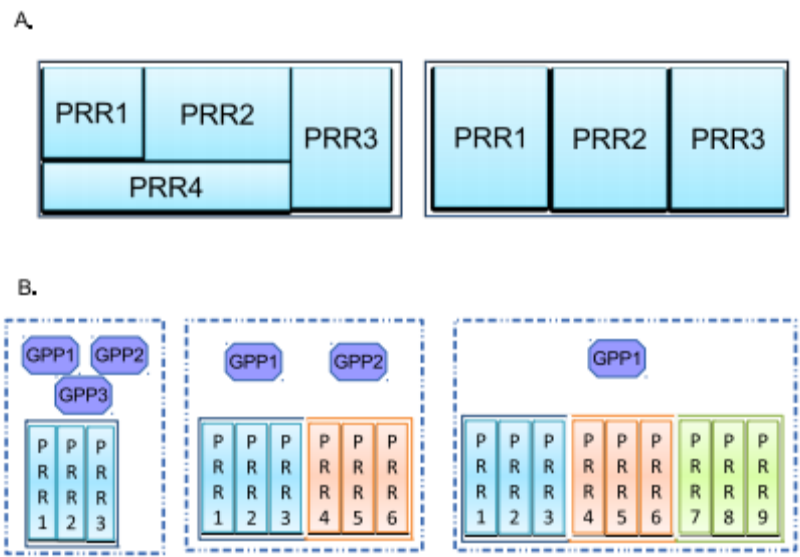

Figure 1. Floorplan decisions: (A) Predicting the layout, (B) Predicting the number of PRRs and GPPs

### 1.4. Contribution
The main contributions of this paper can be clearly stated as following:
1. The majority of work on Reconfigurable Computing Systems rely on a static approach to estimate and decide upon the resources required to solve a specific problem. In this work, we instead propose a novel dynamic and adaptive approach.
2. To the best of our knowledge, the use of data-mining and machine-learning techniques has not been proposed by any research group to exploit this specific type of Design Exploration for Reconfigurable Systems in terms of predicting the appropriate floorplan of an application.

### 1.5. Paper Organization
The remainder of this paper is organized into five sections. Section (2) provides an overview of Reconfigurable Computing and Machine Learning along with necessary background. Section (3) discusses the most significant work published in the field of resource estimation for Reconfigurable Systems. In Section (4) the proposed technique for estimating and predicting the necessary resources is described. Section (5) provides our results based on the proposed implementation. Conclusions and future directions are finally presented in Section (6).

## 2. BACKGROUND
Reconfigurable computing is capable of mapping hardware tasks onto a finite FPGA fabric while taking into account the dependency among tasks and timing criteria. Therefore, managing the resources becomes essential for any reconfigurable computing platform. In this section, we provide the necessary background for the proposed work in this paper. Accordingly, the concept of run-time reconfiguration and Xilinx flow [2] used in this work will be discussed, followed by resource management in the form of an operating system. The concept of machine learning, data mining and classification will then be introduced as a means to predict necessary resources required by the operating system in the proposed framework.

**2.1. Partial Reconfiguration Flow**

Dynamic partial reconfiguration allows several independent bit streams of configurations to be mapped into the FPGA fabric independently, as one architecture can be selectively swapped out of the chip while another is left executing. Partial reconfiguration provides the capability of keeping certain parts intact in the FPGA while other parts are replaced, similar to memory management in traditional CPUs. A Reconfigurable Computing system traditionally consists of a General Purpose Processor (GPP) and several Reconfigurable Modules that execute dedicated hardware tasks in parallel [3]. A fundamental feature of a Partially Reconfigurable FPGA is that the logic and interconnects are time multiplexed. Therefore, in order for an application to be mapped on to an FPGA, it needs to be divided into independent tasks such that each sub-task can be executed at a different time.

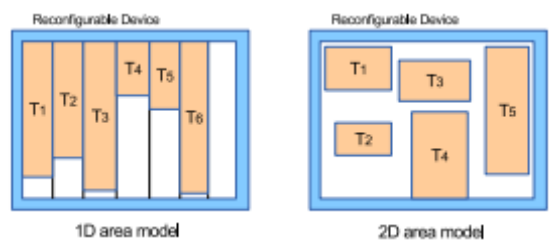

Figure 2. 1D versus 2D task placement area models for reconfigurable devices

Xilinx Partial Reconfiguration design flow [2] uses a bottom up synthesis approach, where Reconfigurable Modules have to be synthesized separately. Each Reconfigurable Module is considered as a separate project where it is verified and synthesized separately. The top design treats each Partial Reconfigurable Region as a black box. After generating all net-lists (top design and Reconfigurable Modules), each Partial Reconfigurable Region must be manually floor-planned using the Xilinx PlanAhead design tool. The PRR can be rectangular or L shaped with some restrictions. More details can be found in [2].

The task placement model in a reconfigurable device can be abstracted as a 1D or 2D model. The 1D model divides the reconfigurable device into columns that can be reconfigured separately, and where a task size is assigned based on width only. In the 2D based approach, a task can have any width and height and can be placed anywhere on the FPGA fabric. The 1D model simplifies the placement mechanism and trades this simplification for a suboptimal device utilization [4]. Both models are shown in Figure 2. Partial reconfiguration is appealing and attractive since it provides flexibility. However, multitasking reconfigurable hardware is complex and requires some overhead in terms of management. In order for users to benefit from the flexibility of such systems, an operating system must be developed to further reduce the complexity of application development by giving the developer a higher level of abstraction. In subsection (2.2.) we will briefly discuss the main components of an operating system for run time reconfiguration platforms.

**2.2. Operating Systems for Reconfigurable Computing**

The use of an operating system for reconfigurable computing should ease application development, help tackle the underlying architecture, and most importantly verify and maintain applications. There are several essential modules that need to exist in any reconfigurable operating system implementation, as shown in Figure 3. The main components of any hardware operating system are the bit-stream manager, scheduler, placer, and communications network.

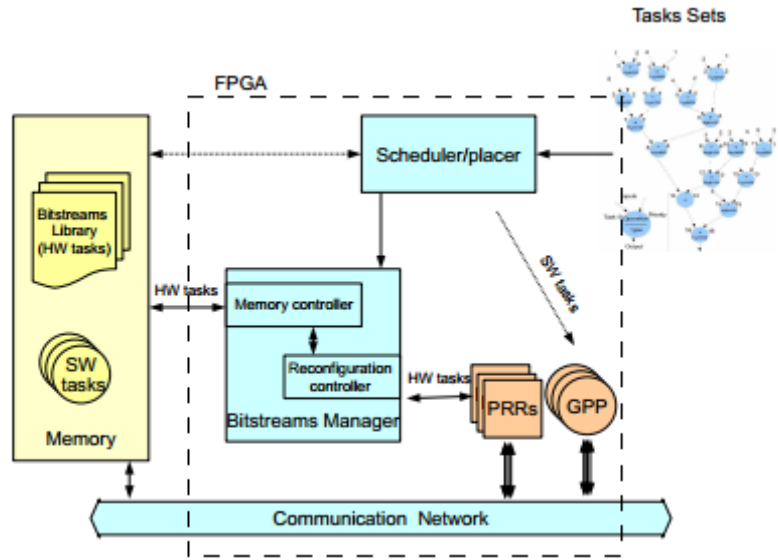

Figure 3. Reconfigurable OS: essential components

1. The Scheduler: In any type of operating system a scheduler usually decides when tasks will be executed. Efficient task scheduling algorithms have to take data communication, task dependencies, and resource utilization of tasks into consideration to optimally exploit the performance of a dynamic reconfigurable system. Schedulers in reconfigurable operating systems differ from conventional software approaches such as Linux or Windows in several ways:
   (a) Reconfigurable operating system schedulers are heavily dependent on the placement of hardware tasks, while in conventional systems the scheduler can be independent from memory allocation.
   (b) Computation resources are available as soon as a task is placed in the reconfigurable fabric, while in conventional systems the task may wait in a ready queue for free processing resources.
   (c) Reconfigurable computing schedulers have to take into account reconfiguration time and should minimize this time by taking into account task pre-fetching and reuse. This is not an issue in conventional operating systems.
   In our previous work several scheduling algorithms were proposed for reconfigurable systems to overcome the limitation of conventional schedulers [5]. Scheduling techniques proposed in the literature have different goals, such as improving hardware resource utilization, reduction of scheduling time and/or reconfiguration overhead. Other reconfigurable computing schedulers attempt to reduce fragmentation and power consumption [6].
2. Bit-stream Manager: This module manages the loading/unloading of partial bit-streams from a storage into Programmable Reconfigurable Regions (PRRs). The bit-stream manager requires fast and fairly large storage media, therefore it is preferable to use a dedicated hardware module for such a task. A bit-stream manager is further decomposed into a storage manager and a reconfigurable manager. The latter attempts to read bit-streams from the storage manager and downloads them onto the FPGA fabric.
3. The Placer: In conventional reconfigurable devices a placer decides where to assign new tasks in the reconfigurable area. In RTR systems the scheduler and placer are inter-dependent. Therefore, they are implemented as a single entity which makes it challenging to identify clear boundaries between them.
4. Communication Manager: This module defines how hardware and software tasks communicate and interact with each other. The communication network proposed by the manager depends on the type of applications used in the system. For example, streaming applications such as vision need a different topology than that of a centralized computational application.
   In current run-time reconfigurable systems, the floorplan and resources are fixed and allocated a priori before the system is deployed. The allocated resources in the form of the number, size, shape of reconfigurable regions and the scheduler type to be used might not be suitable for all types of tasks and applications to be executed on the FPGA fabric. Accordingly, a performance penalty with significant costs could be incurred. Hence, a more suitable dynamic approach is needed for resources estimation and allocation. In subsection (2.3.), we will introduce the main concept of Machine Learning and Data mining to predict and estimate the necessary resources for an application based on past historical information.

**2.3. Data Mining, Machine Learning and Classification**

Data Mining is the core process of the knowledge discovery procedure. A data-mining flow includes different stages, such as Pre-processing, Classification, Clustering and Post-processing. The main objective of the entire process is to extract useful hidden knowledge from the data. Each set of data can be mined using different data-mining techniques depending on the data and the goal of mining procedure, as shown in Figure 4.

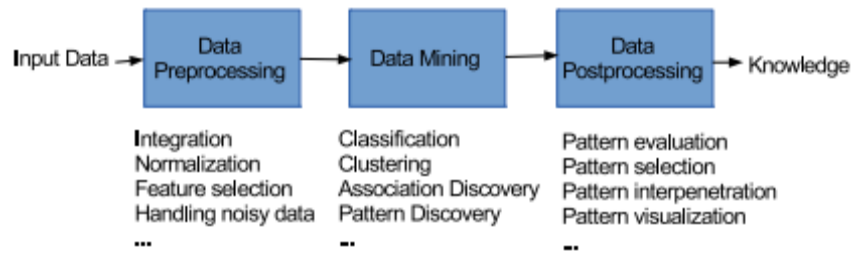

Figure 4. Data mining flow

Classification is one of the main phases of data-mining. It is a key function that categorizes items and records in a database specific classes or categories. The main objective of classification is to accurately predict the target class for each item in the data. Classification is considered to be a form of supervised learning technique that infers a function from labeled data, as shown in Figure 5. The training data usually consists of a set of records. Each record is a pair consisting of an input object along with a desired output target. The training data is analyzed by the supervised learning algorithm and produces a model or function which can be used for mapping new examples. The main objective is to produce a model that correctly determines the class label for unseen instances. In other words, the learning algorithm will have the capability to generalize from the provided training data to unseen situations in some reasonable accurate fashion.

For the work proposed in this paper, a classification model is used to predict the appropriate type of resources and layout for a reconfigurable computing platform given a specific application. Classification in our current work begins with a data set in which the class assignments are known. For example, a classification model that predicts the number of PRRs could be developed based on observed data for many data flow graphs over a certain period of time. Binary classification is considered to be the simplest type of classification problem where the target attribute has only two possible values. For example in our work, the two possible values could be either uniform or non-uniform PRRs. On the other hand, multi-class targets could have multiple values; for example, small, medium or large number of soft cores that can be used in a reconfigurable computing platform.

**3. RELATED WORK**

The use of both machine-learning and data-mining methods, as proposed in this work, represents a new direction for reconfigurable-computing research. In contrast, it has already become a fast-growing research area in physical design. Applications include predicting defects in silicon wafers [7] identifying speed paths in processors as guides for performance improvement [8] and design exploration for high-level synthesis [9]. A notable effort in the area of CAD for ASICs is PADE [10], a new ASIC placement flow which employs machine-learning and datamining methods to predict and evaluate potential data-paths using high-dimensional data from the original design's net-list. PADE achieves 7% to 12% improvements in solution quality compared to the state-of-the-art. A summary of other successful applications of data-mining driven prediction to problems in the area of physical design can be found in [11].

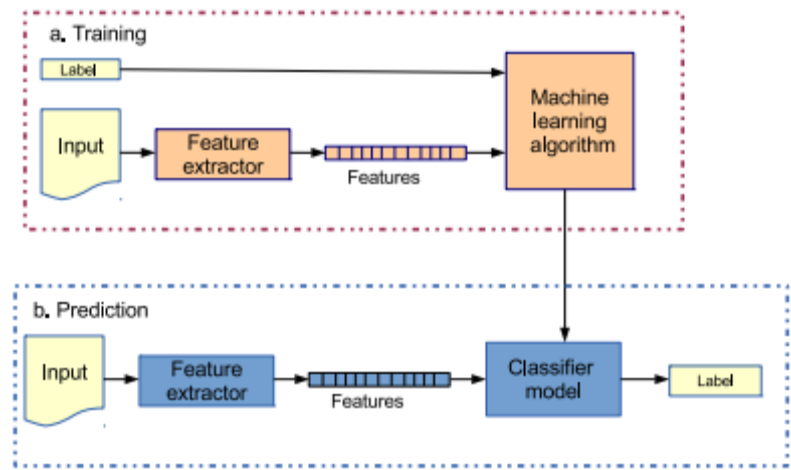

Figure 5. Supervised learning steps

There seems to be an abundance of research work in the literature that covers the concept of management of reconfigurable computing systems. Most of the previous work mainly concentrates on the development of Operating Systems and managers along with the necessary modules such as schedulers and placers. Only very few articles discuss the concept of resource estimation and utilization of machine learning techniques for predicting the necessary resources for dynamic reconfigurable computing systems. [1] presented an automated tool to support dynamic reconfiguration for high performance regular expression searching. The author presented a method to quickly and accurately estimate the resource requirements of a given set of regular expressions. However, this work is limited since it applies only to regular expression searching. Also, no prediction nor learning from past history is applicable in this approach.

In [12] the authors investigate the use of advanced meta-heuristic techniques along with machine learning to automate the optimization of a reconfigurable application parameter set. The approach proposed in [12] is called the Machine Learning Optimizer (MLO), and involves a Particle Swarm Optimization (PSO) methodology along with an underlying surrogate fitness function model based on Support Vector Machines (SVM) and a Gaussian Process (GP). Their approach is mainly used to save time on analysis and application specific tool development. Our work is completely different than MLO in the sense that our framework predicts the necessary resources and floor-plan in Dynamic Reconfigurable Systems to optimize the execution of benchmarks as they arrive for processing.

The authors in [13] propose a fast, a priori estimation of resources during the system level design for FPGAs and ASICs targeting FIR/IFR filters. The prediction was based on Neural Networks. The type of resources they targeted included Area, Maximum Frequency and Dynamic Power Consumption. However, the work is very limited and is not applicable to Dynamic Run Time Reconfiguration applications.

An on-line predictor for a dynamic reconfigurable system is proposed in [14] to reduce reconfiguration overhead by pre-fetching hardware modules. The proposed algorithm uses a piecewise linear predictor to find correlation and load hardware modules a priori. This work tries to optimize the use of fixed resources, while our work plays a role at a much higher level, and seeks to predict the necessary resources.

The work in [15] proposed a dynamic learning data mining technique for failure prediction of high performance computing. The main contribution is to dynamically increase the training set during the system operation, which helps in predicting failures at early deployment. The work in [15] is fundamentally different from our work, since it does not predict resources based on an intelligent machine learning approach.

A multi-objective design space exploration tool is proposed in [16] to enable resource management for the Molen reconfigurable architecture. The proposed approach analyzes an application source code and using heuristic techniques determines a set of hardware/software candidate configuration (sub-tasks). The resource manager then uses these candidates to exploit more efficiently the available system resources. Their work tends to optimize the subtasks of the application to fit a fixed pre-determined platform, while our work predicts a suitable platform for a given application. Their work targeted a specific platform (Molen) and does not take partial reconfiguration into account.

In [17] the authors proposed an algorithm for Programmable Reconfigurable (PR) module generation. The proposed technique can be integrated in manual design flow, to automate the generation of PR partitions and modules. The authors formulated the PR module generation problem as a standard Maximum-Weight Independent Set Problem [18]. Their design supports multiple objectives such as reconfiguration overhead and area with different constraints. This work is different than our work in many aspects. For example, techniques proposed in their paper do not use machine learning nor do they learn from previous results; moreover it is limited to the generation of PR partitions.

The closest published research to our work can be found in [19], [20], [21] and [22]. In [19], the authors presents a high level prediction modeling technique that produces prediction models for is cellaneous platforms and tool chains and application domains. The framework proposed in this paper uses linear regression and neural networks to accurately capture the relation between hardware and software metrics. The framework takes an ANSI-C description as input and estimates various FPGA-related measures, such as area, frequency, or latency. However, our approach is totally different in the following aspects: (a) It does not predict hardware resources consumption but predicts the most optimal layout (PRRs, Soft Cores) that would best execute a certain application such that power is reduced and performance is enhanced, (b) our framework is associated with an Operating System for Dynamic Reconfigurable systems, (c) our framework targets dynamic reconfigurable designs and not static designs as in Quipu.

The work in [20] presents a framework consisting of two layers for resource management of dynamic reconfigurable platforms. The proposed system is capable of evaluating the performance of a reconfigurable computing platform based on prediction model. The framework is applied to an artificial vision case study. However, this paper does not seek to predict neither the layout nor the suitable resources required for the application. The resources are in fact fixed and the main task of the run-time resource manager (RRM) is to allocate the best computational resource (software or hardware) based on the application. The approach in [?, RunTimeOpt-FPL2013]s different since the application level decision making runs a greedy optimization which is computationally expensive to find the best mapping that returns the maximum performance. In our approach we use a smart supervised learning approach to efficiently predict the best layout that would maximize the performance and reduce power consumption.

In [21] the authors propose an online adaptive algorithm that decides the best implementation to be used for the execution of an instance based on features of the process and history of execution. The work tends to improve the hardware/software partitioning task by avoiding predetermined execution times and concentrating on run-time based on system execution history. This work does not use any statistical or machine learning technique to predict resources or floor-plan of the reconfigurable system.

The work in [22] proposes a decision making support framework called DRuid which utilizes machine learning and a meta-heuristic (combines Genetic Algorithms and Random Forest) to extract and learn characteristics that make certain functionality of applications more suitable for a certain computing technology. Starting from a 'C' implementation, the framework either selects the best computational element that can be accelerated by the computational element or offers suggestions on code transformation that can be applied. The expert system identifies 88.9% of the time the functionalities that are efficiently accelerated by using the FPGA. This work, however, is different from our proposed work since it does not predict the most suitable floor-plan nor layout of the reconfigurable computing platform for a specific application, but predicts the functionality of the application that can be accelerated efficiently by an FPGA.

## 4. METHODOLOGY

The ability to determine and predict the hardware resources required by an application in a dynamic, run time reconfigurable environment can significantly improve the system's overall performance in different ways. However, optimizing the necessary hardware resources for a given task graph in real time is an NP-hard problem [23]. Therefore we are proposing a machine learning based technique to estimate and predict the needed resources. Given a new application, our proposed framework employs a database, a simulator and machine learning algorithms to construct a model capable of intelligently estimating the resources required to optimize various objectives such as run-time, area, power consumption and cost. The proposed approach uses previous knowledge and features extracted from benchmarks to create a supervised machine-learning model that is capable of predicting the estimated necessary resources. Figure 6 illustrates the proposed framework and flow used in this work. The framework consists of three main phases: data preparation, training and testing (classification/prediction). The data preparation phase uses both synthetic and real-life benchmarks, a simulator and a database, as shown in Figure 6-A. In this phase, benchmarks are generated and evaluated in terms of the power consumed and execution time using a simulator, as will be explained in Section (4.1.).

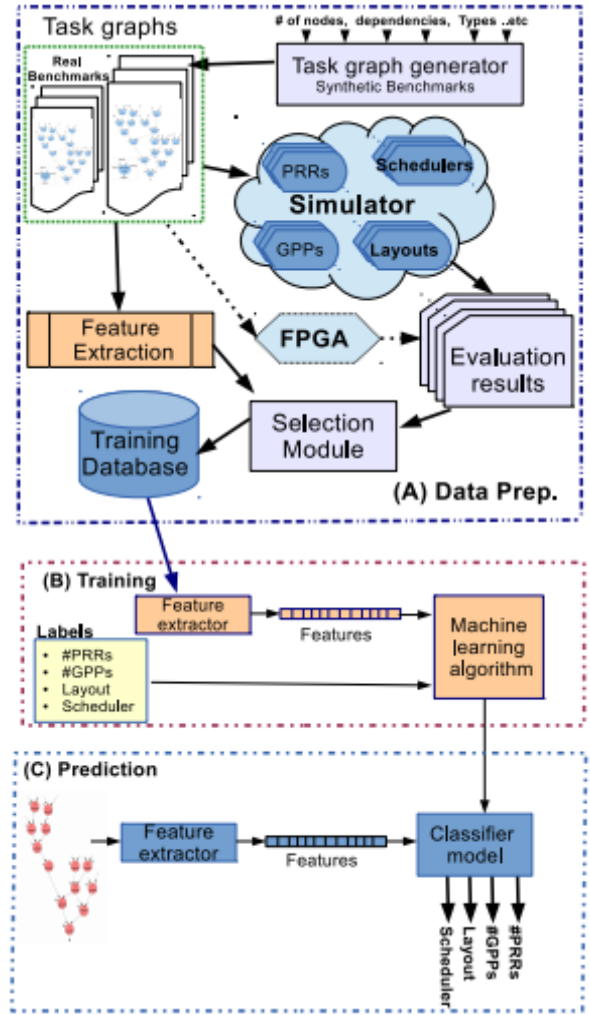

Figure 6. Overall methodology and flow

All necessary features that will be used to train the machine learning algorithms are extracted from both the data flow graphs generated in addition to metrics provided by the simulator. In the training phase (shown in Figure 6-B) the framework utilizes the features extracted from the previous stage to train and create a model that learns from previous historic information. This model extracts useful hidden knowledge (i.e., generalizes) from the data and estimates/predicts resources of the reconfigurable computing system. The third and final testing/prediction phase utilizes the developed model given new unseen task graphs to predict necessary resources as seen in Figure 6-C. Each of these phases are explained in more detail in the following section.

**4.1. Data Preparation Stage**
In this stage a tabular database suitable for the data-mining engine is constructed. Each database record corresponds to a Data-Flow Graph (DFG) along with its features or attributes that are necessary for the training stage. The DFGs can either be synthesized or taken from real-life applications. The synthesized DFGs are evaluated under different hardware set-ups. The evaluation can either be performed using a dedicated hardware platform or a simulator. The latter allows for faster evaluation and flexibility. Each DFG is evaluated using different hardware scenarios, by varying the number of processing elements (PRRs, GPPs), size/shape of PRRs and schedulers. The system performance metrics (power consumption, execution time) for each case is then recorded accordingly. Several features from each DFG such as number of nodes, dependencies, fan-out, critical path, slack, sharable resources and many more are extracted (see Table 2). These features are treated as individual measurable attributes that are effectively exploited in a supervised learning task such as classification.

The necessary modules to construct the database are:

1. Task Graph Generator: The first step in our methodology involves using a task-graph generator to automatically synthesize a large number of input tasks. These input tasks are used later to train and test the proposed predictors. Each input task is represented as a DFG, where the nodes in the DFG represent particular operations to be scheduled and assigned either to hardware or software. The edges on the other hand represent the normal dependencies and flow between operations.

   The data-flow graphs are generated using a similar approach to that proposed in [24]. Each DFG has randomly generated parameters, as shown in Table 1. The probability distribution of DFG parameters is currently based on a uniform distribution, but any other distribution, including a distribution based on real-world applications, can be employed. A total of 258 data-flow graphs are generated at this step in the flow.

Table 1. Data Flow Graphs: Statistics

| Parameter | Range | Note |
|---|---|---|
| Nodes | 5 - 1000 | # of nodes in a DFG |
| Edges | 0 - 1863 | # of edges (dependencies) |
| Edges per Node | 0 - 2 | Average # of edges per node |
| Subgraphs | 1 - 968 | # of Subgrahs in the task |
| Task types | 3 - 16 | # of task types |
| HW tasks area | – | Avg/Min/Max/.. |

2. Simulator (Evaluation): The evaluation of the performance of all DFGs based on different hardware configurations is determined before the features and attributes of each DFG are stored in the database. In our previous work [5] a complete hardware reconfigurable system was designed, mapped and evaluated based on a Xilinx Virtex-6 FPGA board. This platform was initially used for evaluating DFGs. However, using such a platform to evaluate hundreds of DFGs based on different hardware configurations is both complex and tedious. The FPGA platform to be used in this work requires a different floor-plan and bit-stream for each new configuration which limited the scope of testing and evaluation.

   Accordingly a architecture reconfigurable simulator was developed to simulate the hardware platform discussed in [5], while running the developed reconfigurable operating system.

   This simulator utilizes three schedulers and supports any number of PRRs and/or GPPs. The simulator is available under the open source GPL license on GitHub [25]. The simulator is developed to emulate the hardware platform and expects the following configuration files as input:

   - A Task Library file which stores task information used by the simulator. Task information includes the mode of operation (software, hardware or hybrid), execution time, area, reconfiguration time, reconfiguration power and dynamic power consumption (Hybrid tasks can migrate between hardware and software).
     Some of these values are based on analytic models found in [26], [27] and, while others are measured manually from actual implementations on a Xilinx Virtex-6 platform.
   - A Layout file which specifies the FPGA floor-plan. The layout includes the size, shape, and number of PRRs along with types and number of GPPs. The FPGA fabric is first partitioned uniformly then partitioned with 50% increase in size, while varying the number of PRRs. This results in different layouts for the same number of PRRs.
   - A DFG file which stores the data flow graphs to be scheduled and executed.
   - A Configuration file which stores the general settings for the simulator such as the scheduler to be used, resources locations and other parameters.

   In order to have a diverse database we evaluated every DFG with various hardware settings and then recorded the evaluation (performance) metrics. The two most important evaluation metrics used were total execution time and total power consumption for a given DFG. Each DFG was evaluated with different PRR numbers and PRR sizes, various number of GPPs, and schedulers. This resulted in database containing on average several hundreds of records per DFG.

3. Schedulers: Three on-line scheduling algorithms were developed that can handle both hardware and software tasks. The algorithms have different task reuse capabilities to minimize reconfiguration overhead. Tasks are capable of migrating between software (GPP) and hardware (PRRs) to minimize total execution time and mitigate resource scarcity issues. The schedulers record system metrics, learn, and accommodate future tasks.

   The schedulers where first developed and implemented on a hardware platform, and then incorporated into the simulator that is shown in Figure 6. The schedulers are described in more details in [5].

4. Feature Extraction: Since the DFG format cannot be used directly by the classification tool Waikato Environment for Knowledge Analysis (WEKA) [28], a software module was designed and implemented to extract numerous numeric features from a given DFG and store them in a tabular format. Some of the challenges were to extract as much numeric information as possible without adding irrelevant information (noise). Features and attributes were represented either as a single numeric entity or a range format. Below are some of the key extracted features:
   - DFG Connectivity: number of edges, average edges per node, fan-in/fanout, maximum number of parent nodes and maximum number of dependent nodes.
   - DFG size: number of sub-graphs, maximum number of nodes in sub-graph, length of the critical path, daverage length of critical path of the sub-graph.
   - Task types: the number of task types, the number of tasks of each task type, number of hardware/software/hybrid tasks.
   - Schedule flexibility: node mobilities and reuse possibilities.
   - Task type metrics: hardware tasks area, latency, reconfiguration time, and power consumption.
   
   A complete list of the features is shown in Table 2.

### 4.2. Training Stage

The second stage of the framework is related to training and model development, as shown in Figure 6-B. In a prediction problem, a model is usually given a data-set of known data on which training is performed (training data-set), and a data-set of unknown data (or first time seen data) against which the model is tested (testing data-set). In the training phase, a data-set is used to train the machine-learning algorithm so that a model is developed, which can be used in the actual prediction and classification phase. The generated data entries from the previous phase consist of hundreds of records for each DFG that differ in the evaluation metrics such as power and speed. These records were obtained by evaluating the same DFG using different hardware configurations. Since the DFG features are the only metrics that will be used to predict the class (resource type), having multiple identical records per DFG, yet with different classes will have a negative effect on the learning capability of the classifiers. In this step, the fittest record is selected for each DFG (to eliminate duplication). The term fittest could refer to speed, power, cost, or a combination (depending on the designer's goal). For example if the goal is to train the classifier to predict a hardware configuration for the lowest execution time, the fittest function will only keep the records (for each DFG) that has the lowest execution time. Accordingly, several smaller databases are created, one for each objective (power, speed, cost, execution time vs. cost). Each database contains records equivalent to the number of DFGs available.

Cross validation is an important task in machine learning. It is mainly used to estimate how precise and accurately a classification model will perform in real life. One of the main goals of cross validation is to avoid problems such as over-fitting [29] (memorizing) and give a perception of how well the model will generalize to independent data sets. Several techniques for cross validation are available such as (i) the holdout out method, (ii) K-fold cross validation, (iii) Leave-one-out cross validation. In the holdout method, the data set is separated into two different sets called the training set and testing set. The main advantage of the holdout method is that it takes a short time to compute. The second cross validation method "Leave one out" has a good validation error but is very expensive. In our current work, we resorted to the K-fold cross validation as it represents an improvement over the "Holdout method" and is not as expensive as the "Leave One Out" method. The data set in the K-fold cross validation is usually divided into k subsets, and the holdout method is repeated k times. In each experiment, one of the k subsets is used as the testing set while the remaining k-1 subsets form a training set. The average error across all k trials is then computed.

Table 2. Extracted Features from DFGs (These features were used for the training and predication phases)

| Feature | Category Range | Note |
|---|---|---|
| Nodes | 5 - 1000 | # of nodes in a DFG |
| Root Nodes | 0-254 | # of Root nodes without a parents |
| Internal nodes | 0-267 | # of Nodes that are neither root nor leaf. |
| Leaf nodes | 0-502 | # Of leaf Nodes |
| Isolated nodes | 0-940 | # of Nodes with no parent nor children |
| Edges | 0-1863 | # of edges (dependencies) |
| Edges per Node | 0-2 | Average # of edges per node |
| Max parents | 0-2 | Maximum # of parents for a child node. |
| Max Children | 0-19 | |
| Sharable resources | 0-19 | Sum of number of sharable resources |
| Subgraphs | 1-968 | # of Subgrahs in the task |
| Critical paths | – | # Avg, Min critical paths |
| Critical path | – | count the longest path |
| Task types | 3-16 | # of task types |
| Task type N | – | Frequency of each task type |
| HW/SW task types | – | # of HW or SW tasks |
| Migratable tasks | – | # of migratable tasks. |
| Slack | 0-563 | Average slack |
| HW/SW latency | – | Avg/Min/Max values |
| HW/SW execution*pwr | – | Avg/Min/Max values |
| HW config time and power | – | Avg/Min/Max values |
| HW tasks area | – | Avg/Min/Max values |

**4.3. Classification Stage**

The result of the training phase is a model capable of classifying and predicting resources as shown in Figure 6-C. Several supervised machine learning algorithms are deployed and contrasted by measuring the accuracy of prediction. Each classifier is executed with their default parameter values, then evaluated using the test sample for each DFG group. Finally the mean accuracy of the validations is calculated. Classifiers used in this work range from simple Naive Bayes [29] to more accurate and complex machine learning algorithms in the form of ANNs and SVMs [30]. The classification algorithms are trained and developed using the training database using WEKA. A detailed evaluation of the classifiers is discussed in the next section. Since ensemble based systems provide favorable results compared to single expert machine learning systems under certain scenarios, it is worth considering them in this work. Researchers from various disciplines consider and resort to ensemble based classifiers whenever they seek to improve prediction performance. The goal of the ensemble based method is to create a more accurate predictive model by integrating multiple models. The ensemble methodology attempts to weigh and combine several individual classifiers in order to obtain a unified classifier that outperforms each individual classifier and is similar to human beings seeking several opinions before making any important decision. Accordingly, to further improve the performance of the proposed framework, ensemble based classifiers [31] are employed. An important feature of the proposed methodology is that the system is able to learn, generalize and, therefore, can be expected to improve its accuracy over time. The goal of the entire process is to extract useful hidden knowledge from the data. This knowledge is the prediction and estimation of the necessary resources for an unknown or not previously seen application.

**5.   EXPERIMENTS AND RESULTS**

The primary objective of this work is to demonstrate that our proposed system is able to generate accurate prediction models, suitable for use in reconfigurable operating systems, based on known features of data-flow graphs extracted from previous designs. To achieve this goal, five different cases were developed and evaluated. The first two (I-II) are based on predicting schedulers (described in [5]) that can minimize power consumption, migrate hardware/software tasks, and minimize total execution time. The remaining three cases (II-IV) are based on predicting the number of soft cores (GPPs) and partial-reconfigurable regions (PRRs) to be used on the reconfigurable platform to minimize combinations of total execution time, power, and area. The cases that will be discussed are:

1. Predict the type of scheduler that will lead to a minimum power implementation while allowing one or more GPPs to be employed in the design.
2  Predict which of the three scheduler will result in a minimum power schedule while only allowing only hardware tasks (PRRs) to be used in the design.
3. Predict the number and type of the PRRs and GPPs required to minimize execution time.

4. Predict the number and type of PRRs and GPPs required to achieve a balance between power consumption and execution time. (Notice that this case involves multiple, possibly competing objectives.)
5. Predicts the number of the PRRs necessary to minimize both execution time and the total area consumed by the design.

Table 3 summarizes the cases along with the classes that will be predicted. Note that the last case is based on three different layout of PRRs.

Table 3. Summary of Five Cases used to Evaluate the Framework along with Specifications

| Case # | Class type | # of Classes | Objective | Notes |
|---|---|---|---|---|
| I | Scheduler | 2 | Min Power Consumption | For GPPs and PRRs |
| II | Scheduler | 2 | Min Power Consumption | No GPPs (PRR only) |
| III | GPPs/PRRs | 4 | Min Execution Time | - |
| IV | GPPs/PRRs | 4 | Min Execution Time/Power Consumption | - |
| V | PRR | 3 | Min Execution Time and FPGA fabric (Cost) | Figure 1-B |

It is important to note that the results we are achieving and highlighting in this section are not limited to the previous five cases. The framework and methodology proposed is more general and flexible enough to enable the development of prediction models for any combination of resources related to the application or underlying run-time environment including communication infrastructure, interfaces, etc.

**5.1. Classification Algorithms: Implementation**

Machine-learning algorithms proposed in the literature can vary in complexity, accuracy and performance. They can range from simple to more complex models based also on the parameters that need to be tuned which usually influence the training procedure. Each algorithm has its own advantages and disadvantages, and no one algorithm works best in all cases. Consequently, in this work, we will first attempt to use the following five individual classification algorithms, all of which are different in terms of complexity and accuracy, to implement and evaluate each of the five previous prediction cases (i.e., Case I-V). In Sec. 5.5. we will resort to more advanced classification techniques (Ensemble of Classifiers) that combine several individual classification algorithms to further improve the prediction performance and accuracy.

- Naive Bayes (NB): Based on Bayes theorem, an NB classifier classifies a record by first computing its probability of belonging to each class, and then assigning the record to the class with the highest probability [32].
  Advantages of this classifier include its simplicity, computational efficiency, and classification performance. However, for good performance, the classifier requires a large number of records and assumes that the individual feature values extracted from the records are independent, which is not always the case.
- Multi-Layer Perceptron (MLP): Based on a model of biological activity in the brain, MLP is a feed-forward neural network that has proved to be among the most effective models for prediction in the context of data mining [33]. Although able to capture highly complicated relationships between the predictors and a response, their flexibility and performance relies heavily on having sufficient data for training purposes.
- Support Vector Machine (SVM): Related to statistical learning theory, SVM is a non-probabilistic supervised learning algorithm where training records are mapped into a higher dimensional input space so that the separate classes are divided by an optimal separating hyperplane (i.e., a hyperplane with maximum separation margin) in this space [7]. New input records are then mapped into this space and predicted to belong to the class based upon which side of the hyperplane they fall. Although directly applicable to binary-class problems, SVM can still perform effectively in high-dimensional spaces. However, this often requires the overhead of solving a series of binary classification problems.
- K-Nearest Neighbor (K-NN): The K-NN method is both simple and lacks the parameter tuning of many of the methods described above [34]. It is a lazy learning algorithm since it defers the decision to generalize beyond the training examples till a new query is encountered. It works by first identifying the k-nearest neighbors in the training set to the new record. The new record is then assigned to the predominant class among these neighbors. Despite its simplicity, the K-NN method performs remarkably well, especially when the target classes are characterized by a number of related features.
- J48: Tree methods, like J48, are considered to be among the most robust methods for performing classification and prediction. In general, these methods work by recursively splitting the training data into subgroups based on the data's features. In the case of J48, the decision to split is based on the concept of information entropy and seeks to choose the features that maximize the normalized information gain [7]. The recursive splitting stops and leaf nodes are created, upon finding subsets belonging to the same class.

The classifiers listed above were used to classify and predict each of the five cases presented earlier in Section (5.). Thus a total of 25 models were developed and verified. Each of the 25 models was implemented in Java.

### 5.2. Experimental Setup and Evaluation

Each of the 25 models was first trained and tested using 258 synthetic benchmarks (data-flow graphs) described in Section (4.1.). Following that, all 25 models were tested using 20 real-world benchmarks from Media-bench DSP benchmark suite [35]. All of the models were evaluated based on their accuracy. The accuracy of a prediction model is a measure of how well the predictor is able to make correct predictions, and is formally defined as the ratio of correctly classified instances to the number of total instances, as shown in Equation 1.

$$Accuracy = \frac{T_N + T_P}{T_N + T_P + F_P + F_N} \qquad (1)$$

Note: $T_P$ and $T_N$ are the number of true positives and true negatives, respectively, while $F_P$ and $F_N$ are the number of false positives and false negatives, respectively. In context of this paper, an accuracy of 1.0 would mean that the model was able to predict the necessary resources (i.e, scheduler, numbers and types of GPPs and/or PRRs) to achieve an optimal objective (e.g., minimum execution time, power, and/or area) 100% of the time. In addition, we also evaluate each model using the well-known Receiver Operating Characteristics (ROC) curve [36–38]. The ROC curve is a plot of the $T_P$ rate verses $F_P$ rate, and shows the trade-off between sensitivity and specificity. The ROC evaluation is more appropriate measure than accuracy when dealing with imbalanced data sets, as will be explained later. In the following subsections, the detailed experimental results are discussed.

### 5.3. Results for Synthetic Benchmarks

The methodology used to evaluate the performance of the 25 prediction models was based on 10-fold crossvalidation. All 258 data-flow graphs described in Section (4.1.) were randomly divided into 10 subgroups, each containing an equal number of data-flow-graphs. Each of the 25 prediction models was then trained using 9 of the subgroups and then tested using the tenth subgroup. The process was repeated 10 times and the average accuracy for each model was computed. The accuracies of the prediction models were then compared using a t-test (with alpha = 0.05) corrected to avoid Type I errors due to the dependence between samples.

Based on Figure 7 and Tables 4 and 5, the results conclusively demonstrate at a high rate of statistical significance that J48, SVM and MLP achieve the highest accuracy rate of around 83%. This means that 83% of the time the prediction model is able to predict the necessary resources to optimal objective. The other classification models, based on K-NN and NB, obtain a lower accuracy rate around 79%.

Interestingly, the graph in Figure 7 also shows that the average accuracy of all five classification methods decreases as the number of classes increases. In particular, the average accuracy of all five classification methods for case I and case II, which represent binary classification problems, are 93.7% and 94.2%, respectively. For case V, which is a 3-class problem, the average accuracy of all five classification methods is 84.3%. Finally, the worst average accuracy (62%) occurs for case III, where the number of classes increases to 4. This behavior may be attributed to imbalanced data among the various problem instances. The class imbalance problem occurs when the distribution of class instances is skewed between classes. An unbalanced distribution causes typical classification algorithms, which are designed to maximize classification accuracy, to have trouble learning the minority class or classes. As can be seen from Figure 8, the class imbalance problem is more noticeable in the multi-class data sets than in the binary-class data sets. The problem of imbalanced data can be dealt with using either undersampling or oversampling which might not be ideal if the dataset size is either small or duplicating data may effect the models. A more conservative and promising approach used by many researchers to reduce the impact of within-class imbalance is to utilize ensemblebased methods, as explained in the next subsection.

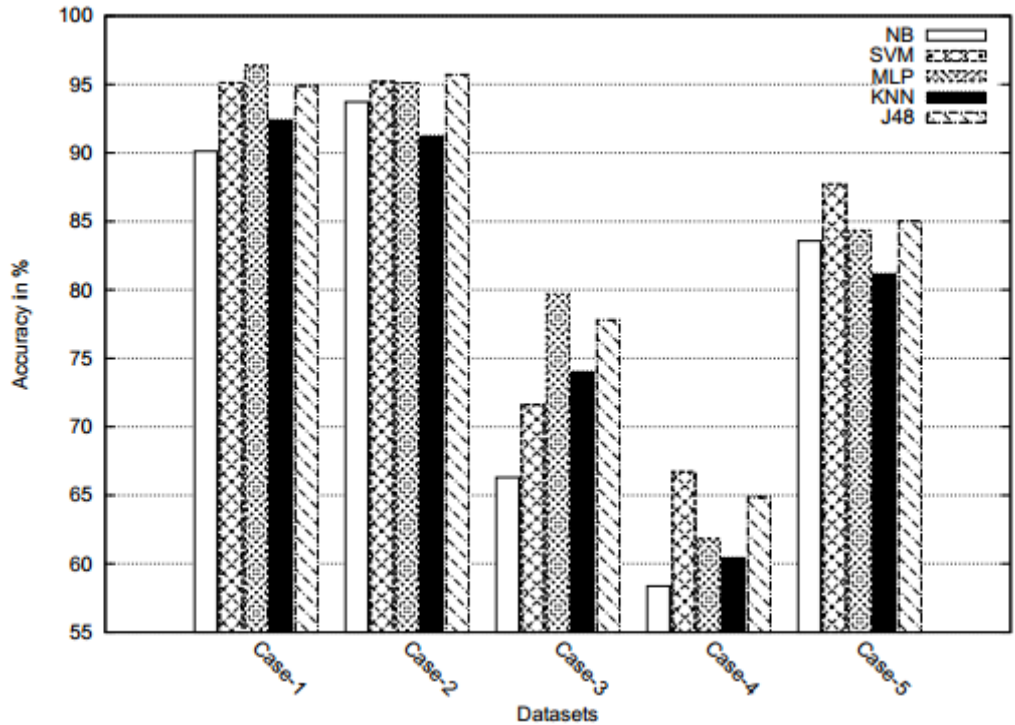

Figure 7. Individual classifiers: predication accuracy for the synthesized benchmarks

Table 4. Accuracy with T-test significant evaluation for the SVM algorithm [synthesized benchmark]

| Dataset | SVM | NB | MLP | KNN | J48 |
|---|---|---|---|---|---|
| Case-1 | 95.12 | 90.12 • | 96.36 | 92.45 | 94.89 |
| Case-2 | 95.19 | 93.72 | 95.11 | 91.27 • | 95.71 |
| Case-3 | 71.61 | 66.30 • | 79.69 ◦ | 74.06 | 77.80 ◦ |
| Case-4 | 66.73 | 58.37 • | 61.79 | 60.45 • | 64.85 |
| Case-5 | 87.72 | 83.57• | 84.34 | 79.89 | 85.01 |
| Average | 83.27 | 78.42 | 83.46 | 81.20 • | 83.65 |

◦, • statistically significant improvement or degradation

Table 5. Accuracy with T-test significant evaluation for the J48 algorithm for [synthesized benchmark]

| Dataset | J48 | NB | SVM | MLP | KNN |
|---|---|---|---|---|---|
| Case-1 | 94.89 | 90.12 • | 95.12 | 96.36 | 92.45 |
| Case-2 | 95.71 | 93.72 | 95.19 | 95.11 | 91.27 • |
| Case-3 | 77.80 | 66.30 • | 71.61 • | 79.69 | 74.06 |
| Case-4 | 64.85 | 58.37 | 66.73 | 61.79 | 60.45 |
| Case-5 | 85.01 | 83.57 | 87.72 | 84.34 | 81.20 |

◦, • statistically significant improvement or degradation

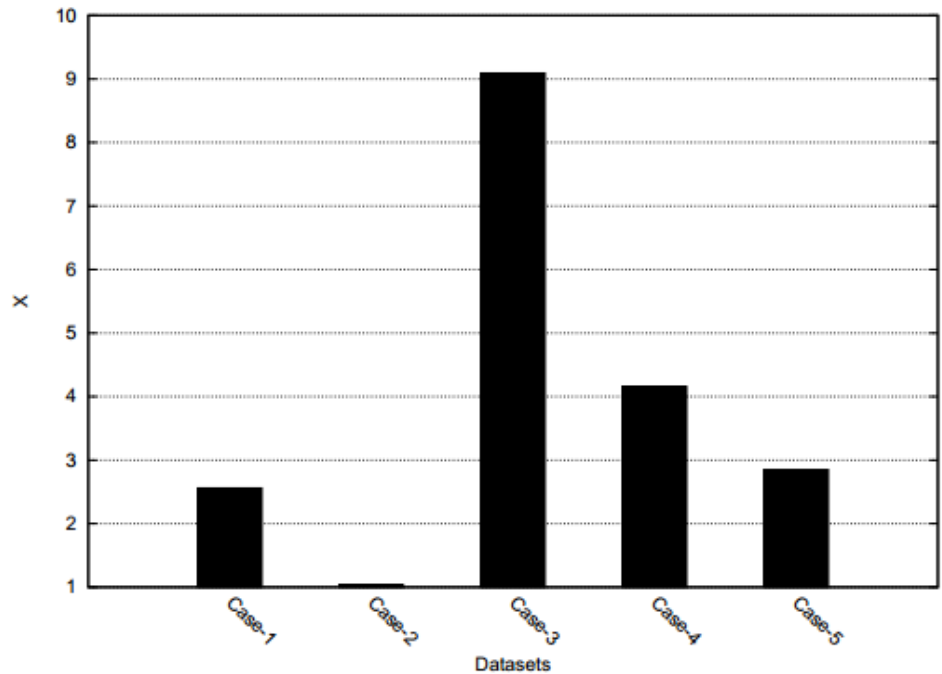

Figure 8. Imbalance ratio between minority/majority classes, where X represent the number of folds (i.e. 2X is 200%) [synthesized benchmark]

With regards to run-time, the training of each classier incurs a one-time computational cost. Table 6 shows the average run-times for training the prediction models based on NB, MLP, SVM, K-NN, and J48, respectively. (Note: each java-based model ran on an Intel Xeon (W3670) workstation with 16 GB of memory.) For example, the SVM algorithm is trained at a one-time cost of around 200 mSec. The Naive Bayes comes handy since it can be trained very quickly compared to SVM, J48 ad MLP algorithms. It should be noted that in the case of K-NN, there is no initial training. Rather, the computational cost is incurred when performing the prediction on the new data-flow graph.

Table 6. Training time for Case # 1 for individual classifiers

| Classifier | Training Time |
|---|---|
| SVM | 100 mSec |
| MLP | 60.4 Sec |
| NB | 10 mSec |
| KNN | - |
| J48 | 200 mSec |

### 5.4. Results for Media Bench DSP suite

In addition to the synthetic benchmarks used in the previous subsection, 20 real-world DFGs based on DSP applications were used for evaluating the performance of the prediction models. The DFGs were selected from MediaBench a standard DSP benchmark suite. Table 7 lists the characteristics of the DFGs that range in complexity from 18 to 359 nodes and from 16 to 380 edges. Figure 9 shows a similar trend to that of the synthetic data in Figure 7 with all of the classifiers performing better on cases 1, 2, 5, 4 and 3 in that order. On average MLP performed the best with average accuracy of 75% followed by SVM and J48 (Average accuracy of 73% and 72% respectively). NB was the worst with average accuracy of 68% while KNN has average accuracy of 70%. For Case-IV J48 performed the best with 57% accuracy. The prediction of the real-word data was lower than the synthetic by an average of 10%.

Table 7. Mediabench DSP Benchmark Specifications

| ID | Name | # of nodes | # of edges | Avg. Edges per node | Critical path length | parallelism (node/critical path) |
|---|---|---|---|---|---|---|
| 1 | JPEG - Write BMP Header | 106 | 88 | 0.83 | 7 | 15.14 |
| 2 | JPEG - Smooth Downsample | 51 | 52 | 1.02 | 16 | 3.19 |
| 3 | JPEG - Forward Discrete Cosine Transform | 134 | 169 | 1.26 | 13 | 10.3 |
| 4 | JPEG - Inverse Discrete Cosine Transform | 122 | 162 | 1.33 | 14 | 8.71 |
| 5 | MPEG - Inverse Discrete Cosine Transform | 114 | 164 | 0.44 | 16 | 7.125 |
| 6 | MPEG - Motion Vectors | 32 | 29 | 0.91 | 6 | 5.33 |
| 7 | EPIC - Collapse pyr | 56 | 73 | 1.3 | 7 | 8 |
| 8 | MESA - Invert Matrix | 333 | 354 | 1.06 | 11 | 30.27 |
| 9 | MESA - Smooth Triangle | 197 | 196 | 0.99 | 11 | 17.9 |
| 10 | MESA - Horner Bezier | 18 | 16 | 0.89 | 8 | 2.25 |
| 11 | MESA - Interpolate Aux | 108 | 104 | 0.96 | 8 | 13.5 |
| 12 | MESA - Matrix Multiplication | 109 | 116 | 1.06 | 9 | 12.11 |
| 13 | MESA - Feedback Points | 53 | 50 | 0.94 | 7 | 7.57 |
| 14 | HAL | 11 | 8 | 0.72 | 4 | 2.75 |
| 15 | Finite Input Response Filter 11/h2 | 44 | 43 | 0.98 | 11 | 4 |
| 16 | Finite Input Response Filter 2 | 40 | 39 | 0.975 | 11 | 3.64 |
| 17 | Elliptic Wave Filter | 34 | 47 | 1.38 | 14 | 2.43 |
| 18 | Auto Regression Filter | 28 | 30 | 1.07 | 8 | 3.5 |
| 19 | Cosine 1 | 66 | 76 | 1.15 | 8 | 8.25 |
| 20 | Cosine 2 | 82 | 91 | 1.11 | 8 | 10.25 |

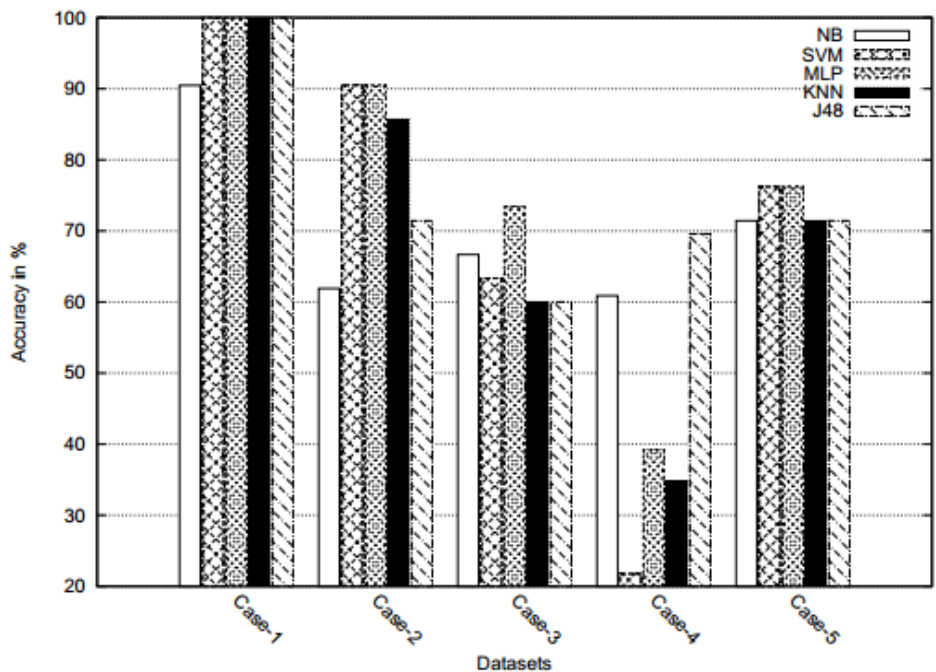

Figure 9. Individual classifiers: predication accuracy for the mediabench DSP benchmark

Except of case-I all cases have higher accuracy in synthetic data. The difference was in the ranged from 9% difference for Case-II to 21% difference for Case-IV. While Case-II and Case-V were in between by an average of accuracy difference of 14% and 11% respectively. The difference in accuracy between the synthetic and the MediaBench DSP benchmark is not surprising. The reason is the models were trained and tested with synthetic data in the first case, but trained with synthetic data and tested with real-world data in the second. This shows that although our models can handle circuits it has not seen before, for best results it would be best to train the model on real-world data.

## 5.5. Ensemble Learning

In this subsection we seek to determine if the accuracy of the prediction models can be improved by employing them in an ensemble methods. Ensemble methods [31] employ not one, but multiple classifiers to classify new data points, often by taking a weighted average vote of their predictions. Ensemble learning algorithms are primarily used to improve prediction accuracy by avoiding problems like variance and bias, or over-fitting when the data set is small. Many techniques have been proposed for combining the predictions of multiple classifiers, but the most popular methods are:

- Bagging is a method that seeks to train each classifier in the ensemble using a random redistribution of the original data. If X is the size of the original training set, training sets are generated for each classifier by randomly selecting X records, with replacement. The predictions of each trained classifier are combined using a (weighted) voting scheme. In practice, bagging is especially useful in situations where even small changes in the training set may lead to big changes in the prediction [39].
- Boosting is an ensemble technique where learning models are learned sequentially. Initial prediction models tend to be very simple, and are used to determine particular data points that are difficult or hard to fit. Later prediction models focus primarily on those points that are hard to fit, with the goal of trying to predict them correctly. In the end, all of the models are given weights and the set is combined to evolve an increasingly more complex and accurate prediction model [40].
- Stacking applies different types of classification models to the original data. Instead of using voting or a weighted average approach, stacking uses a meta-lever classifier to determine the winning model. In general, stacking works well, but it is very hard to analyze [7].
- Randomization (Random Forest) works with a large collection of decision trees. The method works by generating a large set (forest) of independent decision trees by using different random samples of the original data set. Voting is then used to determine the final class [7].

In general, each of the previous ensemble methods seeks to create a more accurate classifier by combining less-accurate ones. Figure 10 and Figure 11 compare the prediction accuracy of the single classifier, J48, to that of the four ensemble methods Random Forrest, Boosting, Bagging, and Stacking for both synthetic and MediaBench DSP benchmark respectively. (Note that J48 was selected for comparison, since it achieved the highest average accuracy rate - 83% along with MLP and SVM. It is clear that in every case all four ensembles achieve higher average accuracy rates than J48. The improvements in average accuracy range from 1% to 11%. However, there is no statistically significant difference in average accuracy between the four ensembles.

As observed in the case of the individual classifiers, the average accuracy of all four ensembles decreases as the number of classes increases. In particular, the average accuracy of all four ensembles for case I and case II, which represent binary classification problems, are 96.15% and 96.03%, respectively. For case V, which is a 3-class problem, the average accuracy of the ensembles is 86.16%. Finally, the worst average accuracy (70.9%) occurs for case III, where the number of classes increases to 4. Although the use of ensembles does not fully solve the problem of class imbalance, it does help to mitigate its effects. Notice that the largest improvements in average accuracy compared with J48 occurred in case 3 and case 4 both of which have more than 2 classes.

Table 8 shows the average accuracy achieved by all five individual classifiers and all four ensembles. It is clear that ensembles outperform all of the single classifier approaches. Table 9 shows the average training time for each of the classifiers. Although there is no statistically significant difference between the four ensembles in terms of accuracy, both bagging and stacking require an order-of-magnitude more training time compared with random forest and boosting. This would suggest that the latter two techniques should be used due to their run-time efficiencies.

The area under ROC for the ensembles techniques is shown in Figure 12 and Figure 13 for both synthetic and Media-bench DSP suite respectively. It is clear that the average performance of the ensemble based techniques is better than the J48 machine learning technique in all cases.

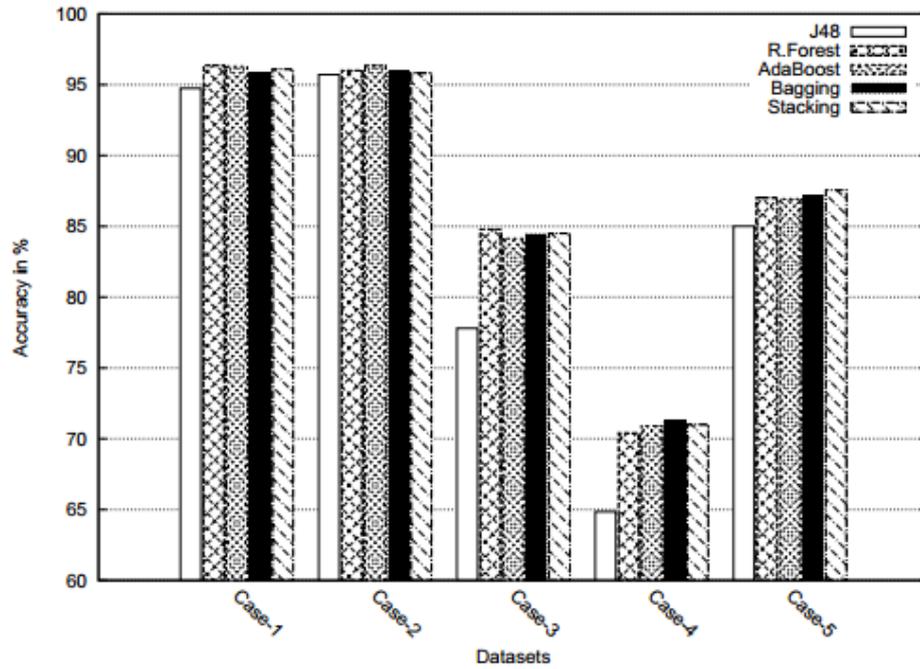

Figure 10. Ensembles classifiers: predication accuracy for the synthesized benchmark

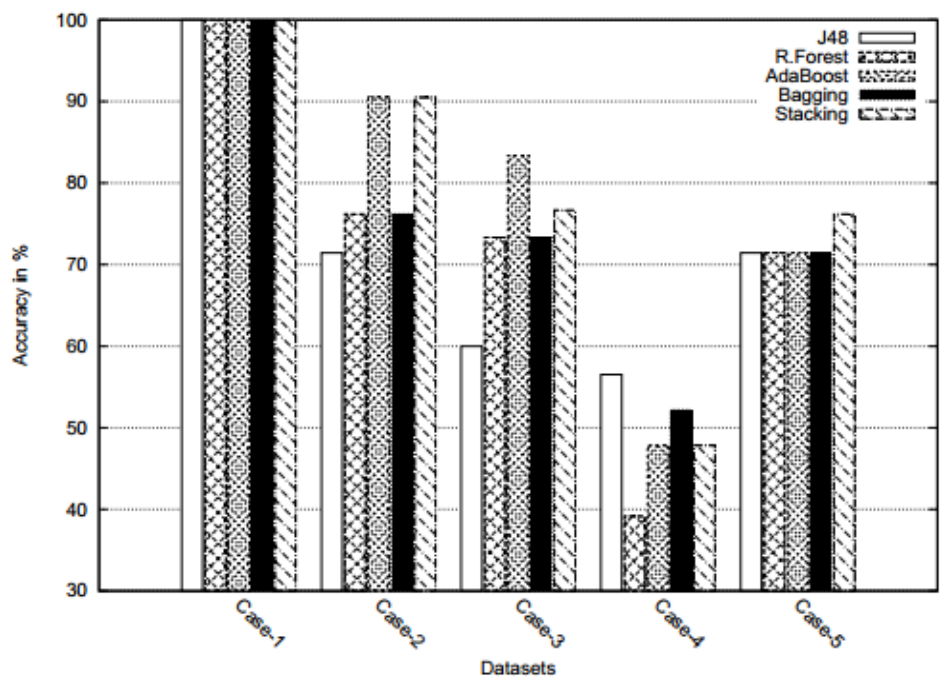

Figure 11. Ensembles classifiers: predication accuracy for the mediabench DSP benchmark

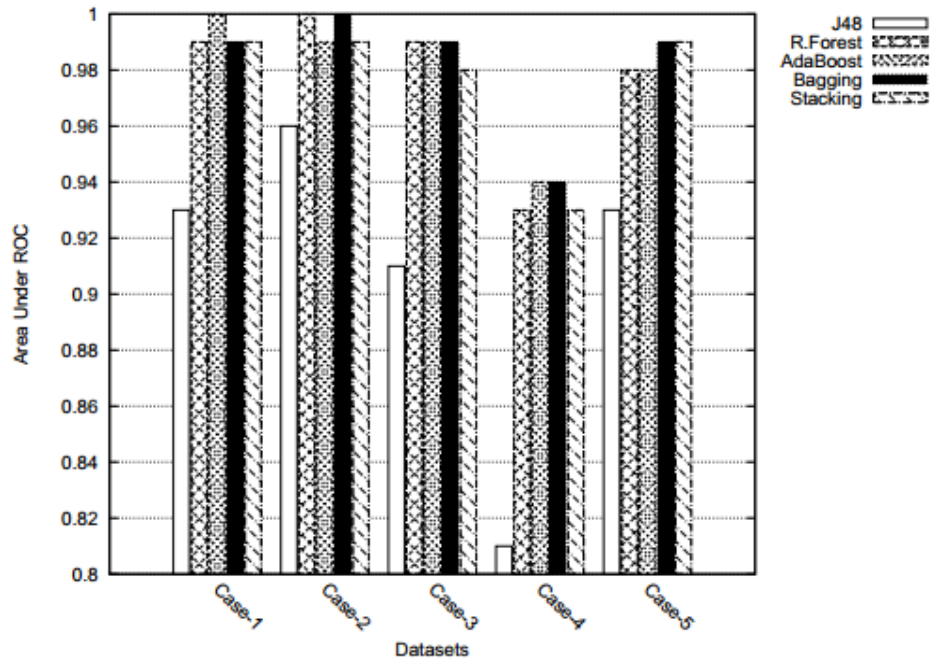

Figure 12. Ensembles classifiers: area under ROC for the synthesized benchmark

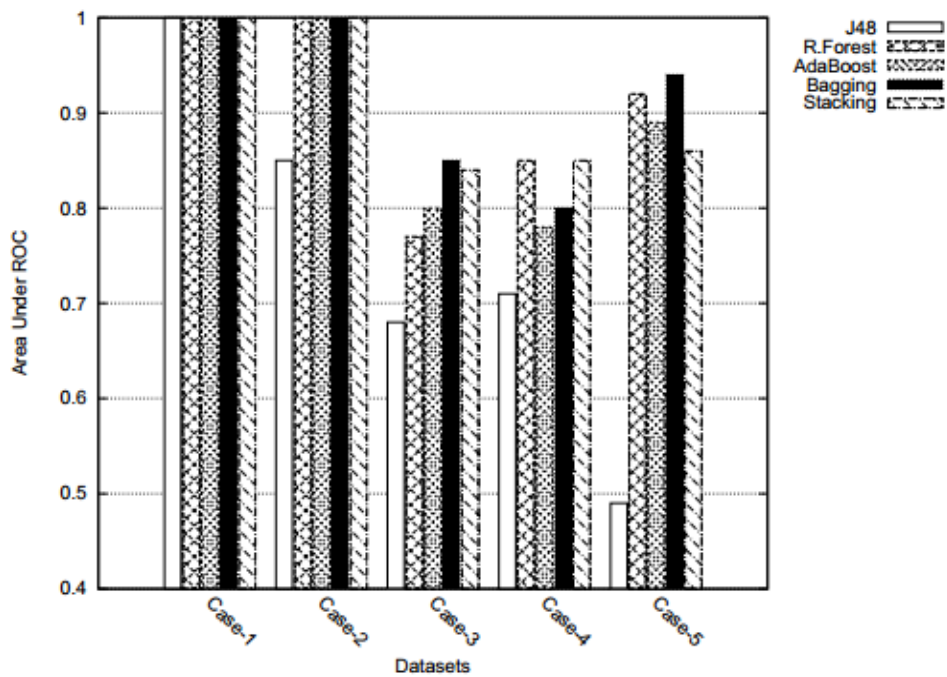

Figure 13. Ensembles classifiers: area under ROC for the media bench DSP benchmark

Table 8. Accuracy with T-Test significant evaluation for the MLP algorithm [synthesized benchmark]

| Dataset | MLP | NB | SVM | KNN | J48 | Rand. Forest | AdaBoost | Bagging | Stacking |
|---|---|---|---|---|---|---|---|---|---|
| Case-1 | 94.74 | 90.12 • | 95.31 | 96.67 ∘ | 92.60 • | 96.36 ∘ | 96.29 ∘ | 95.86 ∘ | 96.09 ∘ |
| Case-2 | 95.71 | 93.72 • | 95.19 | 95.11 | 91.27 • | 96.01 | 96.32 | 95.97 | 95.82 |
| Case-3 | 77.80 | 66.30 • | 71.61 • | 79.69 ∘ | 74.06 • | 84.79 ∘ | 84.14 ∘ | 84.45 ∘ | 84.50 ∘ |
| Case-4 | 64.85 | 58.37 • | 66.73 | 61.79 • | 60.45 • | 70.41 ∘ | 70.90 ∘ | 71.27 ∘ | 71.00 ∘ |
| Case-5 | 85.01 | 83.57 • | 87.72 ∘ | 84.34 | 81.20 • | 87.01 ∘ | 86.94 ∘ | 87.14 ∘ | 87.56 ∘ |

∘, • statistically significant improvement or degradation

Table 9. Training time for different classification algorithms used [synthesized benchmark].

| Dataset | MLP | NB | SVM | KNN | J48 | Rand. Forest | AdaBoost | Bagging | Stacking |
|---|---|---|---|---|---|---|---|---|---|
| Case-1 | 51.90 | 0.01 ◦ | 0.10 ◦ | 0.00 ◦ | 0.10 ◦ | 0.40 ◦ | 0.40 ◦ | 3.60 ◦ | 5.80 ◦ |
| Case-2 | 40.90 | 0.01 ◦ | 0.10 ◦ | 0.00 ◦ | 0.10 ◦ | 0.40 ◦ | 0.40 ◦ | 3.50 ◦ | 5.70 ◦ |
| Case-3 | 83.90 | 0.02 ◦ | 0.50 ◦ | 0.00 ◦ | 0.60 ◦ | 1.60 ◦ | 1.70 ◦ | 14.60 ◦ | 27.40 ◦ |
| Case-4 | 63.30 | 0.02 ◦ | 0.50 ◦ | 0.00 ◦ | 0.40 ◦ | 1.20 ◦ | 1.30 ◦ | 10.70 ◦ | 20.80 ◦ |
| Case-5 | 44.00 | 0.01 ◦ | 0.20 ◦ | 0.00 ◦ | 0.20 ◦ | 0.70 ◦ | 0.70 ◦ | 6.00 ◦ | 10.60 ◦ |
| Average | 56.8 | 0.014 | 0.28 | 0.00 | 0.28 | 0.86 | 0.90 | 7.68 | 14.06 |

◦, • statistically significant improvement or degradation

### 5.6. Random Baseline

As a final demonstration of the efficiency of our approach, we provide a comparison of our prediction models with that of a random predictor. Working with synthetic data first, 11 DFGs were randomly selected (With uniform probability) out of the pool of 258. These DFGs were then compared with DFGs predicted by the models for Case-III and Case-I using SVM classifiers on the bases of minimizing execution time and power. Table 10 shows the results. An asterisk (*) indicates the model missed the best class yet still close enough and much better than the random guess almost all cases. Similarly a baseline comparison were performed for the MediaBench DSP benchmark for Case-III and Case-I but using the J48 classifier. The result are shown in Table 11.

Table 10. Performance Enhancement (Synthetic DFGs): Prediction Engine vs Randomly Selected Architecture

| DFG ID | # nodes | # of Cycles [Case III -SVM] | | | Power mW [Case I - SVM] | | |
|---|---|---|---|---|---|---|---|
| | | Random | Best | ML-RTR | Random | Best | ML-RTR |
| 9 | 100 | 27,610 | 1,317 | *1418 | 1,882 | 1,845 | 1,845 |
| 39 | 75 | 4,464 | 2,226 | 2,226 | 1,550 | 1,419 | 1,419 |
| 40 | 100 | 3,333 | 2,122 | 2484 | 2,166 | 1,755 | 1,755 |
| 43 | 200 | 5,796 | 2,038 | 2,038 | 4,591 | 3,685 | 3,685 |
| 53 | 50 | 3,503 | 1,427 | 1,427 | 940 | 807 | 807 |
| 54 | 75 | 4,816 | 1,685 | 1,685 | 1,695 | 1,420 | 1,420 |
| 61 | 450 | 3,635 | 2,401 | 2,401 | 9,821 | 7,779 | 7,779 |
| 63 | 750 | 8,173 | 3,727 | * 3985 | 17,484 | 13,797 | 13,797 |
| 99 | 1,000 | 4,827 | 4,242 | 4,242 | 27,758 | 24,885 | * 24963 |
| 106 | 125 | 4,988 | 2,207 | 2,207 | 2,814 | 2,319 | 2,319 |
| 112 | 500 | 5,257 | 4,073 | * 5186 | 11,648 | 9,490 | 9,490 |
| Avarage | | 6,946 | 2,480 | 2,900 | 7,486 | 6,291 | 6,298 |

Table 11. Performance Enhancement (MediaBench DSP): Prediction Engine vs Randomly Selected Architecture

| DFG ID | # nodes | # of Cycles [Case III -J48 ] | | | Power mW [Case I - j48] | | |
|---|---|---|---|---|---|---|---|
| | | Random | Best | ML-RTR | Random | Best | ML-RTR |
| 1 | 106 | 2,499 | 1,091 | * 1791 | 2,954 | 2,801 | 2,801 |
| 2 | 51 | 2,143 | 990 | 990 | 1,455 | 1,366 | 1,366 |
| 3 | 134 | 2,794 | 1,995 | 1,995 | 4,257 | 4,065 | 4,065 |
| 4 | 122 | 9,755 | 2,064 | * 2169 | 3,514 | 3,399 | 3,399 |
| 5 | 114 | 13,156 | 2,061 | * 2063 | 3,747 | 3,737 | 3,737 |
| 6 | 32 | 2,774 | 567 | 567 | 781 | 781 | 781 |
| 7 | 56 | 1,265 | 1,061 | 1,061 | 1,416 | 1,373 | 1,373 |
| 8 | 333 | 12,490 | 2,844 | 2,844 | 7,100 | 6,974 | 6,974 |
| 9 | 197 | 13,158 | 2,073 | 2,073 | 5,657 | 5,597 | 5,597 |
| 10 | 18 | 2,567 | 503 | * 679 | 398 | 396 | 396 |
| 11 | 108 | 9,229 | 1,128 | 1,128 | 2,788 | 2,700 | 2,700 |
| 12 | 109 | 3,809 | 1,610 | * 1815 | 2,597 | 2,471 | 2,471 |
| 13 | 53 | 3,642 | 982 | 982 | 1,299 | 1,259 | 1,259 |
| 14 | 11 | 692 | 349 | 349 | 230 | 227 | 227 |
| 15 | 44 | 2,994 | 1,220 | * 1487 | 988 | 978 | 978 |
| 16 | 40 | 1,398 | 1,015 | * 1091 | 1,019 | 990 | 990 |
| 17 | 34 | 1,469 | 799 | 799 | 1,068 | 1,049 | 1,049 |
| 18 | 28 | 983 | 848 | 848 | 697 | 677 | 677 |
| 19 | 66 | 1,696 | 1,507 | 1,507 | 1,319 | 1,253 | 1,253 |
| 20 | 82 | 13,331 | 1,649 | * 1971 | 1,832 | 1,741 | 1,741 |
| Average | | 5,092 | 1,318 | 1,410 | 2,256 | 2,192 | 2,192 |

## 6. CONCLUSIONS AND FUTURE WORK

Performance is one of the fundamental reasons for using Reconfigurable Computing Systems (RCS). By mapping algorithms and applications to hardware, designers can tailor not only the computation components, but also perform data flow optimization to match the algorithm. There are many challenges in adaptive computing and dynamic reconfigurable systems. One of the major understudied challenges is estimating the required resources and predicting an appropriate floorplan for the system. In this paper we propose a novel technique based on machine learning to predict the necessary resources for Dynamic Run Time Reconfigurations. For our work proposed in this paper, a classification model is used to predict the appropriate type of resources for a reconfigurable computing platform given a specific application. Using ensemble based systems provide favorable results compared to single-expert machine learning algorithms under a variety of scenarios. In our future work we propose to not only further enhance the performance and accuracy of the framework developed but also to apply it to predict other important resources such as the communication infrastructure and communication links between soft cores and programmable reconfigurable regions.

## ACKNOWLEDGEMENTS

This work was partially funded by the Natural Sciences and Engineering Research Council of Canada (NSERC). Authors would like to thank the Canadian Microelectronics Corporation for providing us with all the necessary CAD tools (Xilinx, Mentor Graphics) used in this publication.

## DISCLOSURE

The Authors (A. Al-Wattar, S. Areibi, G. Grewal) state that they do not have any personal or financial relationships with the above mentioned commercial identities (Xilinx, and Mentor Graphics) and, accordingly, there is no conflict of interest.